# Maximum Entropy in Turbulence

T.-W. Lee*
*Mechanical and Aerospace Engineering, SEMTE, Arizona State University, Tempe, AZ, 85287*

**Abstract-** Turbulence may appear as a complex process with a multitude of scales and flow patterns, but still obeys simple physical principles such as the conservation of momentum, of energy, and the maximum entropy principle. The latter states that the energy distribution will tend toward the maximum entropy under physical constraints, such as the zero energy at the boundaries and viscous dissipation. For the turbulence energy spectra, a distribution function that maximizes entropy with the physical constraints is a log-normal function, which agrees well with the experimental data over a large range of Reynolds numbers and scales. Also, for channel flows DNS data exhibits an increase in the Shannon's entropy and total viscous dissipation as a function of the Reynolds number in a predictable manner. These concepts are used to determine turbulence energy spectra and the spatial distribution of turbulence kinetic energy, $u'^2$. The latter leads to a solution to the channel flow problem, when used in conjunction with the expression for the Reynolds stress found earlier.

*T.-W. Lee
Mechanical and Aerospace Engineering, SEMTE
Arizona State University
Tempe, AZ 85287-6106
Email: attwl@asu.edu

**INTRODUCTION**

The maximum entropy principle is very useful, in determining the blackbody radiation spectra (Planck, 1901), energy distribution in particles (Cover and Thomas, 1991), and in specifying drop size distributions (Li and Tankin, 1988), as some examples. This principle states that the energy distribution of particles will tend toward the state of maximum entropy under given constraints of the physical system. Turbulence can be considered as a large ensemble of energetic eddies having a spectrum of energy and length scales. Due to the large size of the ensemble, it will come to an equilibrium state of maximum entropy under the constraints of zero energy at the boundary points and total energy content and viscous dissipation. The total energy and the range of length scales that exist in the turbulent flow primarily depend on the Reynolds number. For example, the total turbulence kinetic energy contained in the energy spectrum will be specified by the initial mean velocity and length scale of the flow, in other words by the Reynolds number. In order to find some universal laws concerning the turbulence kinetic energy spectrum, some insights were provided through theoretical analyses (Kolmogorov, 1962; Kraichnan, 1962; Hinze, 1975). The scaling laws from these analyses are plotted, and compared with lognormal distribution and data, in Figure 1. We can see that various scaling laws are tangent to the lognormal distribution in the wavenumber regions of their applicability. Notably, in the inertial subrange the Kolmogorov $k^{-5/3}$ scaling will track the power spectra for a range of length scales that increases with the Reynolds number. We can see in Figure 1 that due to broadening of the energy spectrum with increasing total energy, or Reynolds number, there will be an increase in the wavenumber range where $k^{-5/3}$ scaling will be nearly tangent to the spectrum. Also, the ascending part of the spectra at low wavenumbers (largest eddies) is tracked by $k^4$-scaling (Hinze, 1975), which is tangent to lognormal function at this range. We can see that $k^n$-type of scaling will be tangent to lognormal at least at some point and range of the wavenumber space.

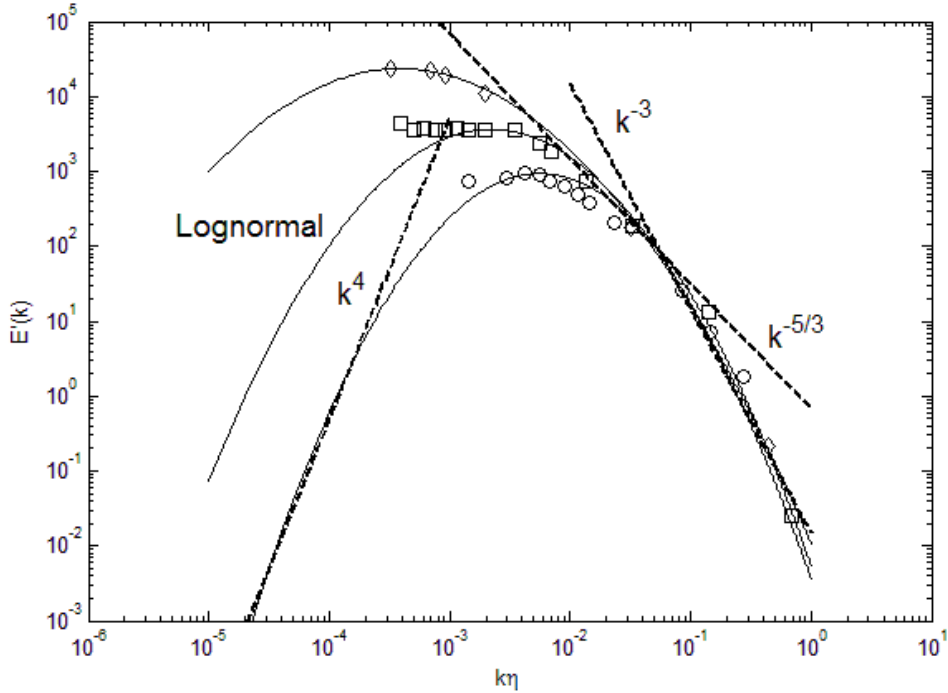

**Figure 1. Log-normal energy spectra and various $k^n$-scaling. The data are from Comte-Bellot and Corrsin, 1975 (circle), Champagne, et al., 1970 (square) and Saddoughi and Veeravalli, 1994 (diamond).**

For turbulence, the maximum entropy principle can be effectively used where the energy spectra will take on a distribution with the maximum entropy, as constrained by the boundary conditions: The kinetic energy must be zero at the smallest dissipation scale (Kolmogorov scale), and also at the largest flow length scale (e.g. dimension of the object in the flow). The range of length scales, or the ratio of length scales from Kolmogorov ($\eta$) to the integral scales ($l_e$), is known from $\eta = l_e Re_\lambda^{3/2}$ (Hinze, 1975), where $Re_\lambda$ is the Reynolds number based on the Taylor microscale (λ). Another important constraint is the total turbulent kinetic energy fixed by the initial or external conditions   The distribution function with the maximum entropy that satisfies these criteria is the lognormal function due to its asymmetric decay to zero at the boundary points. Uniform and exponential distributions have finite boundary values, while Gaussian is symmetric. Energy spectra are asymmetrical because the descent toward zero energy occurs due to physical limit of the flow scale at the low wavenumber extreme, while viscous dissipation causes the approach toward zero at the high wavenumbers. For a similar reason, the drop size distributions in spray flows take on lognormal shape (Lee, 2016). In some cases, the distribution function can be derived algebraically (Bevensee, 1993). Using the Lagrange multiplier method, the distribution function that maximizes the Shannon's entropy $(S = -ElnE)$ for turbulence energy can be derived (see Appendix A), and this form is quite similar to the lognormal function modified $k^2$ terms at the high wavenumbers. An alternate method is to choose the distribution function with the maximum Shannon's entropy that still obeys the physical constraints (Cover and Thomas, 1991), and we have taken this simpler approach. As noted above the width of the distribution can then be deduced from η = $l_e$Re$_\lambda^{3/2}$, while the height of the distribution is set by the total integrated turbulence kinetic energy, which is proportional to the mean velocity squared and the length scales of the flow. For example, atmospheric turbulence will have a very large total integrated energy and also the ratio of largest to the smallest (Kolmogorov) scales will be very large, both of which depend on the Reynolds number.

Lognormality occurs often in nature, including turbulence. Recently, lognormality in turbulence dissipation has been observed in large-scale flows (Pearson and Fox-Kemper, 2018). Intermittency has been modelled as being lognormally distributed in space, but with some conceptual and mathematical inconsistencies that needed to be addressed (Frisch, 1995). Also, Mouri et al. (2008, 2009) have noted on fluctuations of velocity and energy dissipation being lognormal in different flow geometries. However, the maximum entropy principle is directly applicable to energy distributions with intuitive and observable constraint parameters, in this case leading to lognormal turbulence energy spectra.

In this work, we will validate this approach using experimental data over a wide range of Reynolds number. Key attributes of the energy spectra, the height and width, depend primarily on the Reynolds number, and possibly other easily observable parameters. This opens ways to reconstruct the power spectra, as the Reynolds number contains information concerning both the energy and the range of length scales that exist in the flow. There are some secondary parameters that are used in collapsing the power spectra (Saddoughi and Veeravalli, 1994) such as the dissipation and kinematic viscosity, and these may be used to fine-tune the lognormal function. However, the primary parameter for both the height and width of the lognormal spectra is just the Reynolds number, as will be shown.

**MAXIMUM ENTROPY METHOD FOR TURBULENT ENERGY SPECTRA**

As noted above, the distribution function with the maximum entropy that satisfies the physical constraints of turbulent flows and its energy is the lognormal function. At the boundary points, the kinetic energy must be zero at the smallest dissipation scale (Kolmogorov scale), and also at the largest flow length scale (e.g. dimension of the object in the flow). The total turbulent kinetic energy is fixed by the initial or external conditions, and the range of scales from the Kolmogorov ($\eta$) to the integral scales ($l_e$) is known or at least estimable from $\eta = l_e \text{Re}_\lambda^{3/2}$ (Hinze, 1975), where $Re_\lambda$ is the Reynolds number based on Taylor microscale. Energy spectra are asymmetrical because the descent toward zero energy occurs due to physical limit of the flow scale at the low wavenumber extreme, while viscous dissipation causes the approach toward zero at the high wavenumbers. The distribution function with the maximum entropy that satisfy all of these constraints is the lognormal function.

Knowing the total energy and the width of the lognormal distribution allows us to construct the turbulent kinetic energy spectra over a wide range of Reynolds numbers, as shown in Figure 2. We can see in Figure 2 that energy spectra across a very wide range of energy and length scales are accurately determined using the lognormal distribution function (plotted as lines) when compared with data (symbols). Kolmogorov's $k^{-5/3}$ scaling is also plotted (broken line) for comparison, and we can see that for large Reynolds numbers this scaling is tangent to the lognormal distribution in the so-called inertial subrange. This is the region that contains a large portion of the total energy, and thus Kolmogorov scaling has been useful in prescribing the power spectra (Tennekes and Lumley, 1976). Figure 1 and 2 show that quantitatively and qualitatively there is a close agreement between the lognormal distribution and data for turbulent energy spectra over almost the entire length scale range.

It is also interesting to plot the lognormal power spectra in a semi-logarithmic scale as in Figure 3, where there appears to be a shift in the spectra toward higher wavenumber; however, this is only due to nearly 5 orders of energy scales that have been normalized. It is only the wavenumber corresponding to the maximum energy which shifts toward smaller wavenumber as the Reynolds number increases. Also, the energy spectra broadens relative to the Kolmogorov scale ($\eta k$) when the Reynolds number increases (~$Re^{3/2}$). Thus, energy spectra can be specified by knowing the Reynolds number and energy scale of the turbulent flow. We can again see that the energy content is very small toward small length scales (beyond the inertial subrange), which is why $k^{-5/3}$ type of scaling is a good approximation for power spectra.

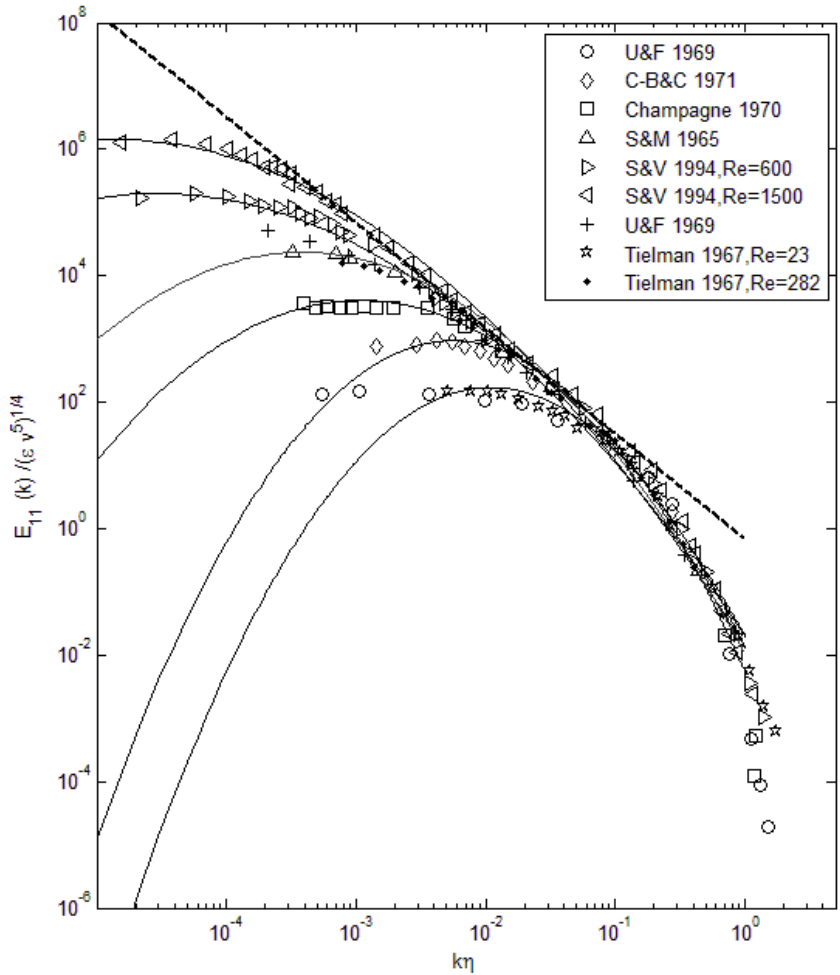

**Figure 2. Log-normal turbulence energy spectra (solid lines) and experimental data. $k^{-5/3}$ fit is plotted as a broken line. Data (symbols) are for one-dimensional power spectra from the work shown in the legend (see references).**

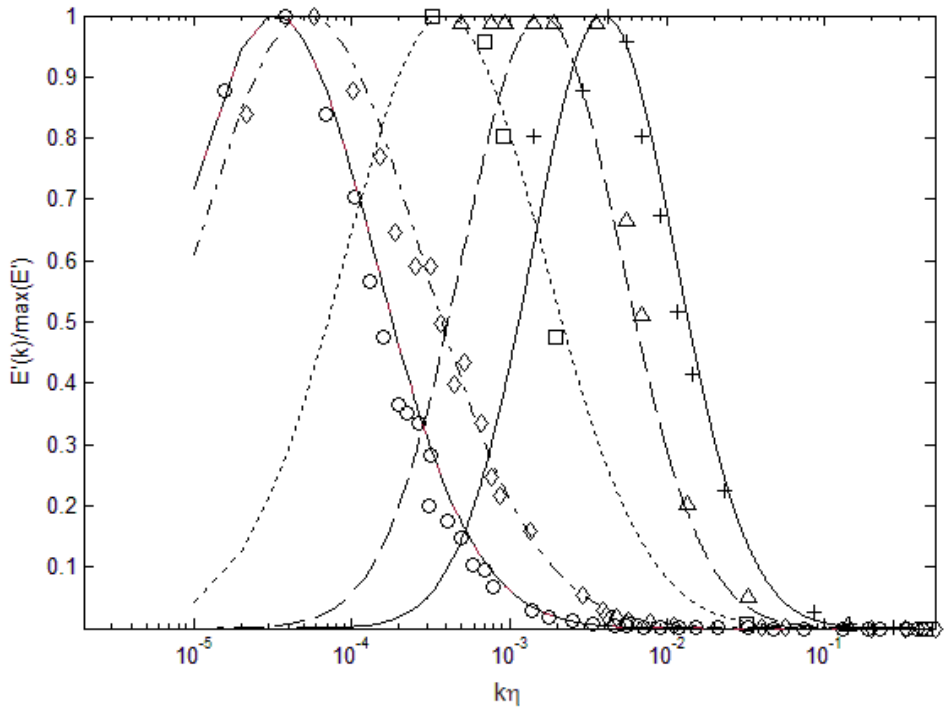

**Figure 3. Lognormal distribution (normalized by the peak value) plotted in a semi-logarithmic scale. The data are from the same references as in Figure 2.**

We can also examine the change in the energy spectra in decaying turbulence. The experimental data of Comte-Bellot and Corrsin (1971) are illustrative of this process, where downstream evolution of power spectra is as in Figure 4. Decay of turbulence energy is exhibited, where both the height and width of the distribution decreases with decreasing local Reynolds number. Yet, the lognormal shape of the power spectra is retained during the decay. Thus, lognormal distributions track both the shape and decaying magnitude of energy spectra. The information concerning the change in the height and width of the spectra is useful, and possibly leads us to some universal functions that can be written in terms of their Reynolds number dependence. In Table 1, we show the maximum energy and width of the spectra, the latter from FWHM (full width at half maximum), both estimated from the data. Note that in Comte-Bellot and Corrsin (1971) the data are given in dimensional units, and due to the sensitivity of the lognormal function to its parameters sometimes this is preferable. Likewise, we can use the data in Figure 2 and extract the height and width information of the energy spectra. These are tabulated in Table 2.

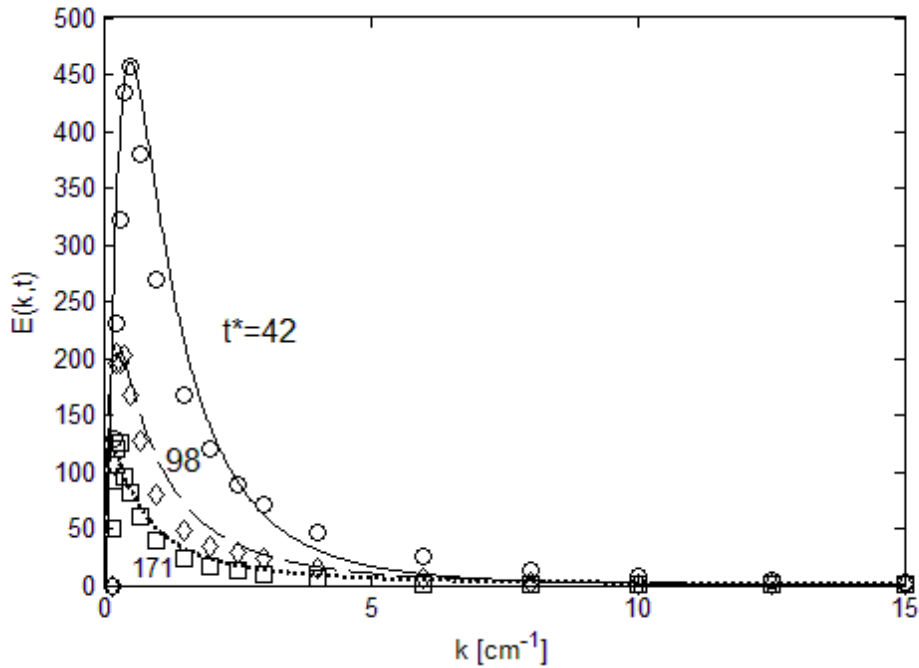

**Figure 4. Temporal decay of energy spectrum. Data are from Comte-Bellot and Corrsin (1971), and t\* normalized time, t\*=tU/M. Local Reynolds numbers are 71.6 (circle), 65.3 (diamond) and 60.7 (square).**

**Table 1.  Parameters of the energy spectra for data of Comte-Bellot and Corrsin (1971).**

| t* | $Re_\lambda$ | $E_{max}$ | $E_{total}$ | $k(E_{max})$ [cm$^{-1}$] | FWHM [cm$^{-1}$] |
|---|---|---|---|---|---|
| 42 | 71.6 | 461.2 | 774.6 | 0.058 | 1.24 |
| 98 | 65.3 | 212.6 | 342.8 | 0.12 | 0.98 |
| 171 | 60.7 | 123.9 | 174.9 | 0.22 | 0.84 |

**Table 2.  Parameters of the energy spectra for various data sets.**

| $Re_\lambda$ | $E_{max}/(\varepsilon\nu^5)^{1/4}$ | $(k\eta)_{Emax}$ | Reference |
|---|---|---|---|
| 37 | 130 | 0.0222 | Comte-Bellot and Corrsin, 1971 |
| 72 | 878 | 0.00439 | Comte-Bellot and Corrsin, 1971 |
| 308 | 53100 | 0.0002084 | Uberoi and Freymuth, 1970 |
| 600 | 199000 | 0.0000578 | Saddoughi and Veeravalli, 1994 |
| 850 | 312900 | 0.000024 | Coantic and Favre, 1974 |
| 1500 | 1446000 | 0.0000179 | Saddoughi and Veeravalli, 1994 |

Using the data in Table 2, we can attempt to find a relationship between the Reynolds number and parameters that go into the lognormal function, so that we can reconstruct the energy spectrum based on the Reynolds number. To adjust the height or the energy scale of the spectrum, we use a multiplicative factor, A, in front of the lognormal function. The lognormal function itself has two parameters, logarithmic mean, $\mu$, and variance, $\sigma$. As shown in Figures 2, 3 and 4, we can see that $\mu$ decreases with increasing $Re_\lambda$, while $\sigma$ increases. Thus, we find a least-square fit to the following functions for these parameters using the data in Table 2.

Pre-exponential factor: $\qquad A = a_1 * Re_\lambda^2 + a_2$ \hfill (1)

Logarithmic mean: $\qquad \mu = b_1 * Re_\lambda^{-3/2} + b_2$ \hfill (2)

Variance: $\qquad \sigma = c_1 * Re + c_2$ \hfill (3)

The form of these functions have been deduced from basic knowledge of turbulence (Hinze, 1975). There may be better functions for these parameters that reconstruct the energy spectra accurately at all the Reynolds numbers and physical configurations. Also, secondary parameters such as $u'^2$, dissipation, kinematic viscosity, and/or other length scales may fine-tune the above functions. However, here we only demonstrate that turbulence energy spectra are recoverable through the log-normal distribution function by using simple function fits for A, $\mu$ and $\sigma$.

An example is shown in Figure 5, where we plot the reconstructed lognormal distributions using the parameters from Eqs. 1-3, and compare with some data. Note that we prefer to use the dimensional wavenumber, k, for this exercise. Energy scale is again $E'(k) = E(k)/(\varepsilon v^5)^{1/4}$.

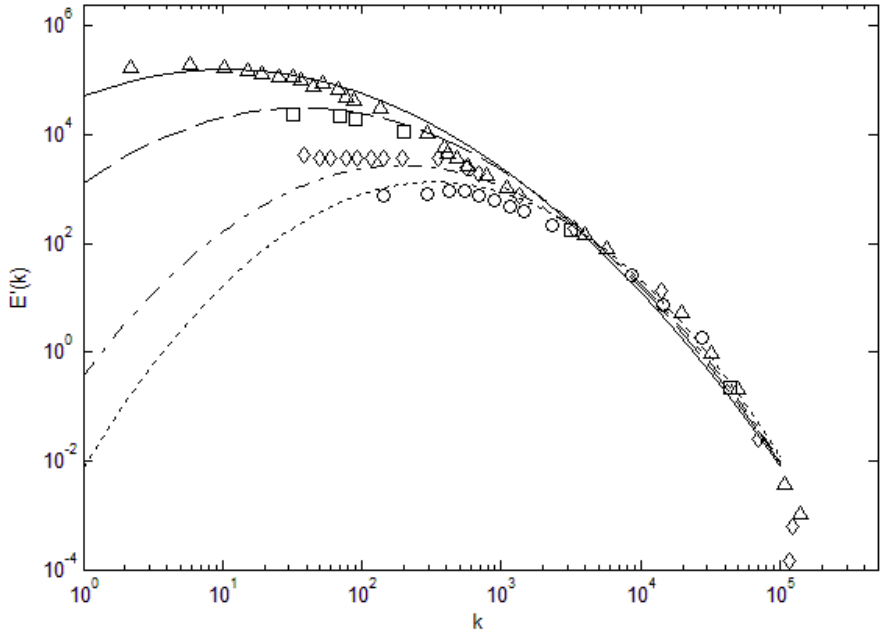

**Fig. 5. An illustration of power spectra reconstruction, using only the Reynolds number as a parameter in Eqs. 1-3. The data are from Comte-Bellot and Corrsin, 1975 (circle), Champagne, et al., 1970 (diamond) and Saddoughi and Veeravalli, 1994 (square), and Sanborn and Marshall, 1965 (triangle)**

The use of the maximum entropy principle is not limited to the wavenumber space. I hypothesize that the spatial distribution of the turbulent kinetic energy will also equilibrate toward the maximum entropy form, while again obeying the physical constraints. For wall-bounded flows, we can observe that spatial distribution of $u'^2$ is highly skewed toward the wall, and at high Reynolds numbers develops a spike, as shown in Figure 2. The descent to zero energy is enforced by the wall boundary condition, while the centerline value is not zero. The other constraints are the total kinetic energy is fixed at a given Reynolds number, and that integrated viscous dissipation increases with increasing Reynolds number. For example, in Figure 3 we plot the total integrated $u'^2$, Shannon's entropy, viscous dissipation and the skewness of the $u'^2$ distributions in Figure 2. When normalized by the friction velocity squared, the total integrated $u'^2$ stays nearly constant meaning that the dimensional u'² increases with the Reynolds number. Viscous dissipation (ε) increases nearly linearly with the Reynolds number, and this is a useful feature in that we can estimate fairly accurately what the total viscous dissipation should be at a given Reynolds number through $\varepsilon Re_\tau$ = constant.

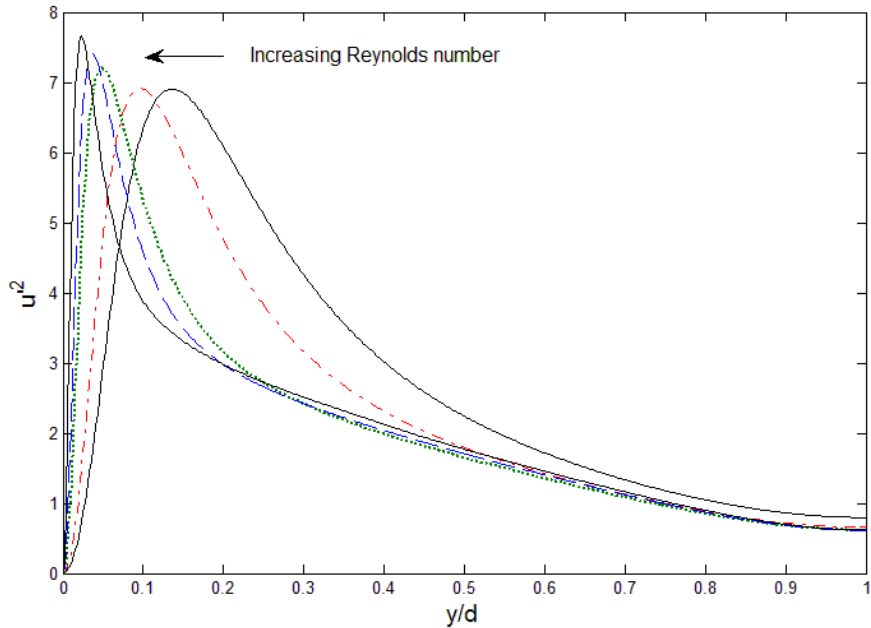

**Fig. 2. u'² profiles from DNS data [12, 13].**

Thus, u'² distribution is of a form that has an asymmetrical distribution that maximizes the entropy production or the viscous dissipation for a given Reynolds number. Observations of the u'² distributions in Figure 2 tend to bear these features, where at very high Reynolds number sharp spikes appear in order to increase the viscous dissipation beyond what is possible through a smooth asymmetrical function such as lognormal function. A combination of functions that satisfies these criteria are lognormal with a Gaussian spike near the wall, as shown in Figure 4. Gaussian spike is more pronounced at high Reynolds numbers since lognormal distribution alone is not sufficient to support high viscous dissipation. The transition from the lognormal to Gaussian is uncertain, and there may be an appropriate function form that smoothly transitions from outer lognormal to inner Gaussian; however, for the current work, we approximate the u'$^2$ distribution as a combination of the lognormal for the outer region and a Gaussian spike for the inner region, as shown in Figure 4. Thus, u'$^2$ distribution will be of the form that matches both the total energy and the viscous dissipation. Looking at the distributions (both observed and reconstructed in Figures 2 and 4, respectively), we can see that the energy balance leading to such a non-linear solution must also be highly non-linear, and we show one example of the governing equation for u'$^2$ in the Appendix B.

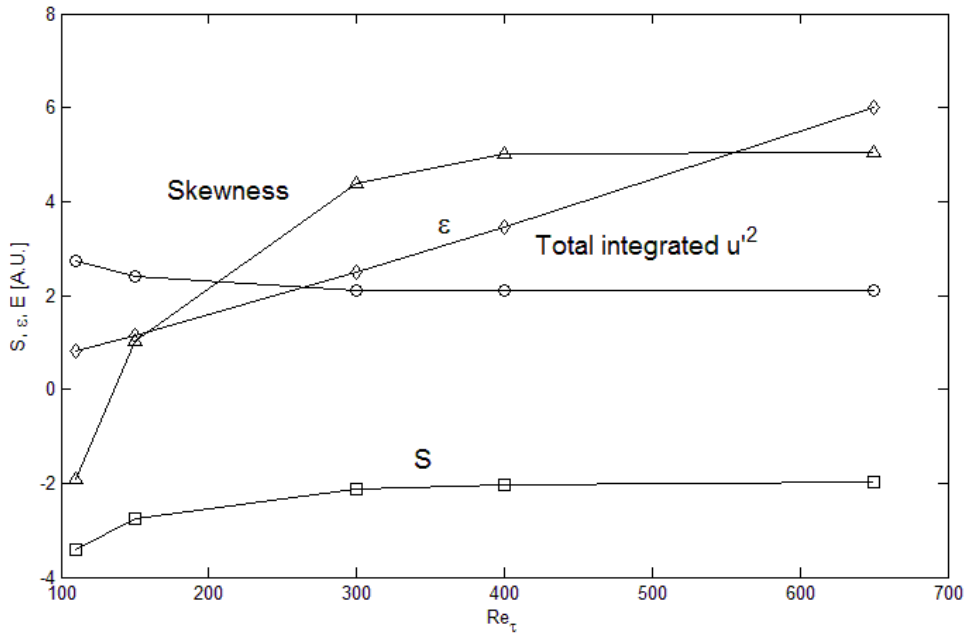

**Figure 3. Shannon's entropy (S), total integrated u'$^2$, viscous dissipation ($\varepsilon$) and skewness as a function of Reynolds numbers from the DNS data [12, 13].**

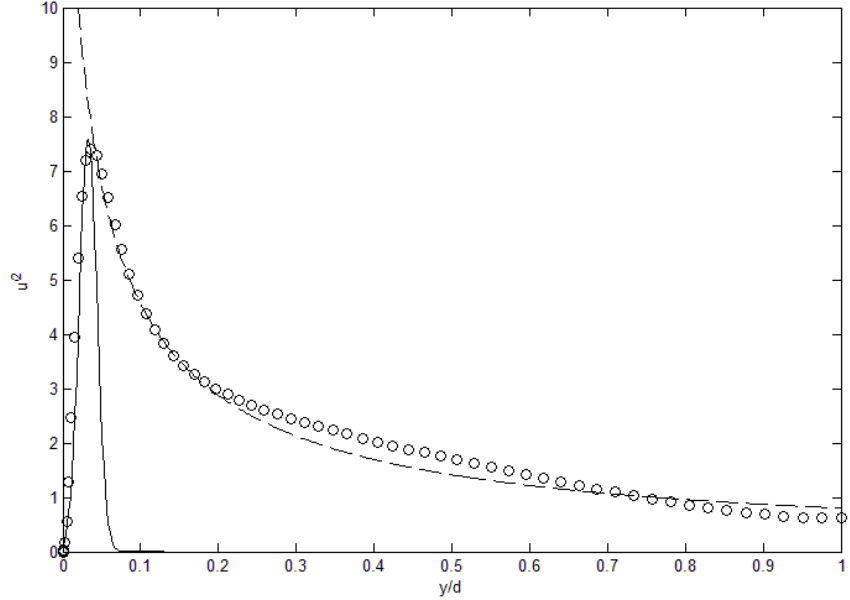

**Fig. 4. u'² profile as a combination of Gaussian (inner) and lognormal (outer, modified to give a finite centerline value at y/d =1). Symbols are the DNS data from Iwamoto et al. [12].**

The shape of u'² profiles suggest that the solution for u'² is probably not so simple to obtain from differential methods, and this leads us to the benefit of using the maximum entropy principle. Even though the lognormal-Gaussian is an approximation for the outer and inner regions, it gives us a short-cut result for the u'² distribution. A recent finding shows that u'² distribution plays a key role in determining the Reynolds stress, where the Reynolds stress gradient is determined from the streamwise transport of u'², pressure force and the viscous term, as shown in Eq. 1 [14, 15].

$$\frac{d(u'v')}{dy} = -C_1 U \left[ \frac{d(u'^2)}{dy} - \frac{1}{\rho}\frac{d|P|}{dy} \right] + \frac{1}{v}\frac{d^2 u'}{dy^2} \tag{1}$$

For channel flows, Eulerian form of the Navier-Stokes equation gives us the mean velocity in the inner and outer regions:

Inner: $\quad \mu \frac{dU}{dy} = \int_0^y \frac{dP}{dx} dy + \rho(u'v')$ \hfill (2b)

Outer: $\quad \mu \frac{dU}{dy} = \rho(u'v')$ \hfill (2b)

For the outer region (Eq. 2b), the momentum transport is dictated by the Reynolds stress [16], while in the inner region (Eq. 2a) the Reynolds stress is relatively small and only modifies the laminar viscous solution. Thus, Eqs. 1 and 2 furnish us with a solvable set of dynamical equations, when combined with the maximum entropy principle (for $u'^2$ distribution). This set of equations can be numerically integrated directly, or found iteratively. For the latter solution, we can start from a plausible mean velocity profile which goes from zero at the wall to the centerline velocity in a continuous, smooth manner, and insert it in Eq. 1 along with the $u'^2$ distribution. Then the velocity profiles can be updated using Eq. 2, until the solution converges. Such a solution for turbulent channel flow is shown in Figure 5. Figure 5(a) shows the Reynolds stress from Eqs. 1 and 2, while the inner and outer mean velocity profiles are plotted in Figure 5(b).

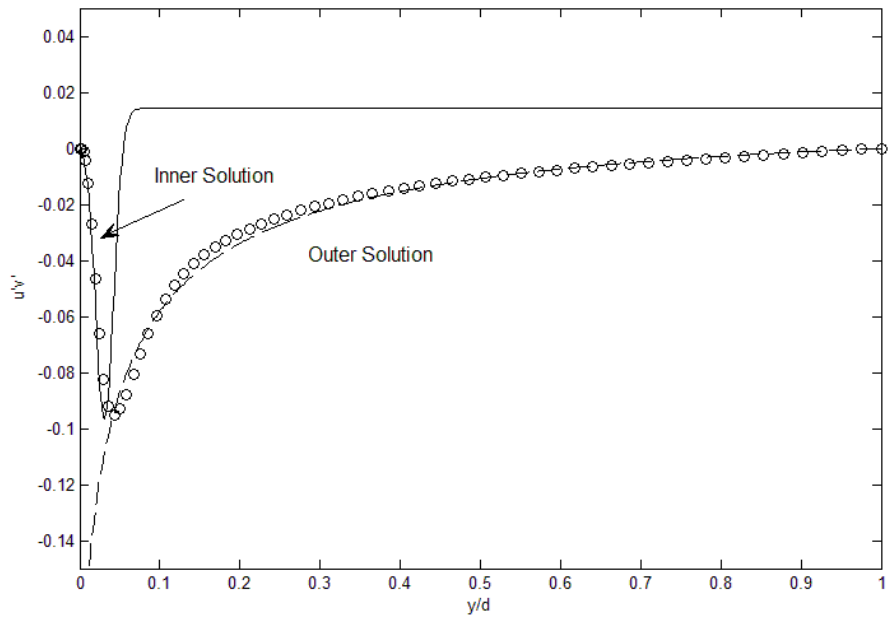

**Figure 5(a). The Reynolds stress profile, compared with DNS data [12, 13].**

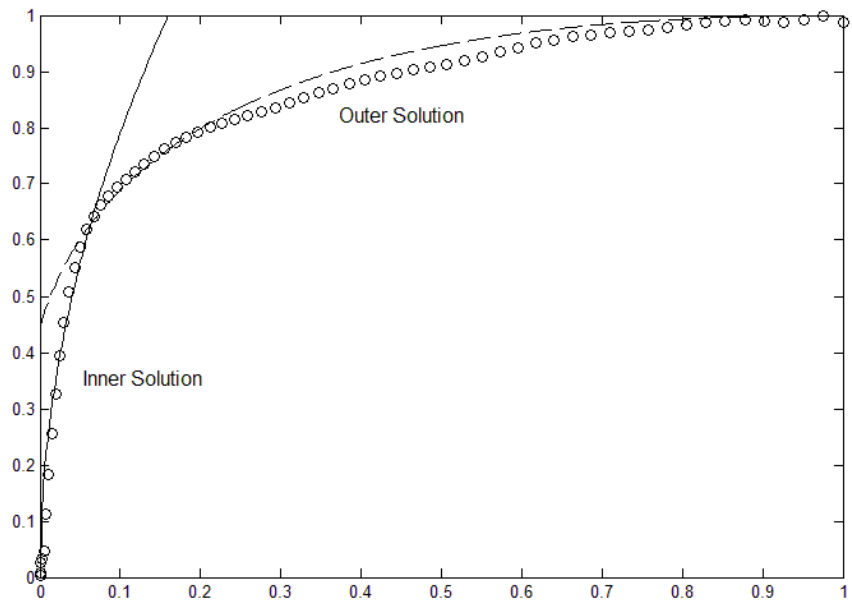

**Figure 5(b). The mean velocity solution, compared with DNS data [12, 13].**

**CONCLUSIONS**

From the maximum entropy principle, we deduce that the lognormal function should prescribe the turbulence energy spectra, as validated with experimental data over a wide range of energy and length scales. The lognormal spectra are consistent with existing scaling laws such as Kolmogorov's $k^{-5/3}$ in the inertial range and $k^4$ dependence in the large-eddy length scales. This approach makes it possible to reconstruct the turbulence energy spectra, using primarily the Reynolds number that determines the width and height of the lognormal distribution. There may be secondary parameters such as dissipation, kinematic viscosity, and lengths scales that can fine-tune the energy distribution, but the fundamental turbulence energy spectra exhibit lognormal behavior that can be prescribed by the Reynolds number, as stipulated by the known properties of the energy and length scales of turbulence.

For the spatial distribution, $u'^2$ profiles are approximated as a combination of Gaussian near the wall and modified lognormal in the outer region. This leads to a solution to the channel flow problem, when used in conjunction with the expression for the Reynolds stress found earlier.

**REFERENCES**

Bevensee, R.M., Maximum entropy solutions to scientific problems, Prentice Hall, 1993.

Champagne, F.H., Friehe, C.A., La Rue, J.C. and Wyngaard, J.C., Flux measurements and fine-scale turbulent measurement in the surface layer over land, J. Atm. Sci., 1977, 34, 515-530.

Coantic, M. and Favre, A., Activities in, and preliminary results of, air-sea interactions research at I.M.S.T, Adv .Geophys., 1974, 18A, 391-405

Comte-Bellot, G. and Corrsin, S., Simple Eulerian time correlation of full- and narrow-band velocity signals in grid-generated isotropic turbulence, J. Fluid Mech., 1971, 48, 2, pp. 273-337.

Cover, T. and Thomas, J. (1991) Elements of Information Theory, John Wiley and Sons, Inc.

Frisch, U., Turbulence, Cambridge University Press, 1995.

Graham, J., Kanov, K, Yang, X.I.A., Lee, M.K., Malaya, N, Lalescu, C.C., Burns, R., Eyink, G, Szalay, A, Moser, R.D. Moser,Meneveau, C., A Web Services-accessible database of turbulent channel flow and its use for testing a new integral wall model for LES, Journal of Turbulence, 2016, 17(2), 181-215.

Hinze, J.O., Turbulence, McGraw-Hill Series in Mechanical Engineering. McGraw-Hill, New York, 1975.

Iwamoto, K. and Nobuhide Kasagi, http://thtlab.jp/DNS/dns_database.htm.

Iwamoto, K., Sasaki, Y., Nobuhide K., Reynolds number effects on wall turbulence: toward effective feedback control, International Journal of Heat and Fluid Flows, 2002, 23, 678-689.

Kolmogorov, N., A refinement of previous hypotheses concerning the local structure of turbulence in a viscous incompressible fluid at high Reynolds number. J. Fluid Mech., 1962, 13, 82-85.

Kraichnan, R.H., The structure of isotropic turbulence at very high Reynolds numbers. J. Fluid Mech. 1959, 5, 497-543.

Lee, T.-W., Quadratic formula for determining the drop size in pressure atomized sprays with and without swirl, Phys. Fluids, 2016, 28, 063302.

Lee, T.-W., and Park, J.E., Integral Formula for Determination of the Reynolds Stress in Canonical Flow Geometries, Progress in Turbulence VII (Eds.: Orlu, R, Talamelli, A, Oberlack, M, and Peinke, J.), pp. 147-152, 2017.

Lee, T.-W., Reynolds stress in turbulent flows from a Lagrangian perspective, Journal of Physics Communications, 2018, 2, 055027.

Li, X. and Tankin, R.S., Derivation of droplet size distribution in sprays by using information theory, Combustion Science and Technology, 1988, 60, pp. 345-357.

Mouri, H., Hori, A., and Takaoka, M., Fluctuations of statistics among subregions of a turbulence velocity field, 2008, Physics of Fluids 20, 035108.

Mouri, H., Hori, A., and Takaoka, M., Large-scale lognormal fluctuations in turbulence velocity fields, Physics of Fluids, 2009, 21, 065107.

Pearson, B. and Fox-Kemper, B., Log-Normal Turbulence Dissipation in Global Ocean Models, Physical Review Letters, 2018, 120, 094501.

Planck, M., Distribution of energy in the spectrum, Ann. Physics, 1901, 4, 3, pp. 553-560.

Saddoughi, S.G. and Veeravalli, S.V., Local isotropy in turbulent boundary layers at high Reynolds numbers, J. Fluid Mech., 1994, 268, pp. 333-372.

Sanborn, V.A. and Marshall, R.D., 1965, Local isotropy in wind tunnel turbulence, Colorado State Univ. Rep. CER 65 UAS-RDM71.

Tieleman, H.W., 1967 Viscous region of turbulent boundary layer. Colorado State Univ. Rep. CER 67-68 HWT21.

Tennekes, H. and Lumley, J.L., First Course in Turbulence, MIT Press, 1976.

Uberoi, M.S. and Freymuth, P., Turbulence energy balance and spectra of the axisymmetric wake, Physics of Fluids, 1970, 13, 2205, doi:10.1063/12693225.

**APPENDIX A**

As noted above, turbulence can be considered as a large ensemble of energetic eddies having a spectrum of energy and length scales. Due to the large size of the ensemble, it will come to an equilibrium state of maximum entropy under the imposed physical constraints: zero energy at the boundary length scales, total energy content and viscous dissipation that are fixed by the Reynolds number. For example, the total turbulence kinetic energy contained in the energy spectrum will be specified by the initial mean velocity and length scale of the flow, and constantly dissipated by the viscous effect, all of which can be parameterized in terms of the Reynolds number. In order to find some universal laws concerning the turbulence kinetic energy spectrum, some insights were provided through theoretical analyses (4-7). We have shown that these scaling laws are subsets of the lognormal spectra [6] that can be obtained from the maximum energy principle.

The energy distribution that maximizes the Shannon's entropy under the physical constraints can be obtained using the Lagrange multiplier method [2, 8]. Here, the principal constraint is that the turbulence conserves energy: the kinetic energy is dissipated by viscosity effect at progressively larger wavenumbers.

$$u'^2 + vk^2 u'^2 \delta t = e_o = constant \tag{1}$$

u'(k) the turbulent fluctuation velocity at a given wavenumber, k, while ν is the kinematic viscosity and δt some time interval. Eq. 1 states that turbulence energy density (on a unit-volume basis) integrated over some time interval δt is conserved. The above constraint can be transposed into the energy distribution using the Lagrange multiplier method (see Ref. 2, for examples).

$$E(k)dV = C_1 exp\{-C_2 u'^2 - C_3 k^2 u'^2\}dV \tag{2}$$

Converting the dV=d(k$^{-3}$) to dk basis, we obtain the following energy distribution.

$$E(k) = \frac{C_1}{k^4} exp\{-C_2 u'^2 - C_3 k^2 u'^2\} \tag{3}$$

In Eq. 3, constants $C_1$, $C_2$, and $C_3$ are determined from the constraints of the turbulence energy content, limiting length scales, and viscosity, respectively. The limiting length scales are the Kolmogorov dissipation length scale and the minimum wavenumber that exists in the flow. We still have the unknown u'(k) in Eq. 3, again to be determined from the physical constraint. In Kolmogorov theory [7], $u' \sim k^{-1/3}$ is obtained in the inertial subrange. However, in general this is an unknown element or a lack of a piece of information. The maximum entropy principle gives the most probable energy distribution under the given physical constraints, but it does not produce unknown information. Thus, the missing pieces of information need to be supplied from observational data, and Eq. 3 provides a framework for inserting the information into the energy distribution.

For example, we can test $u' \sim k^{-n}$ with various n's in Eq. 3 as shown in Figure 1, where it can be seen that this scaling only gives rough trends for the turbulence energy spectra. Kolmogorov spectrum, $k^{-5/3}$ is also plotted in Figure 1, although its applicability is only in the inertial subrange, and the data in Figure 1 is at a relatively low Reynolds number exhibiting little parallel with $k^{-5/3}$ spectrum. An alternate function that has an inverse dependence on k is $u' \sim -\log(k)$. For u' >0, we can use the form u' = m-log(k). Using this form for u', Eq. 3 appears to reconstruct the observed turbulence energy spectrum quite well, with classic maximum at the most probable energy state while decreasing at both ends (high energy density) of the wavenumber space. Comparison with experimental data shows that this form works quite well in replicating the turbulence energy spectra over several orders of magnitude in energy and length scales, as shown in Figure 2. Other constraints are applied to determine $C_1$ (peak energy scale), $C_2$ and m (range of length scales) and $C_3$ (kinematic viscosity) for the theoretical spectra in Figure 2. Note that these constants can easily be parameterized by the Reynolds number [6] and viscosity. When all of these constraints are used in Eq.3, we obtain lognormal distributions modified by viscous dissipation at high wavenumbers. In a previous study, we have used deductive logic to arrive at the lognormal distribution [6]; however, the viscous dissipation effect can now be incorporated into the energy distribution.

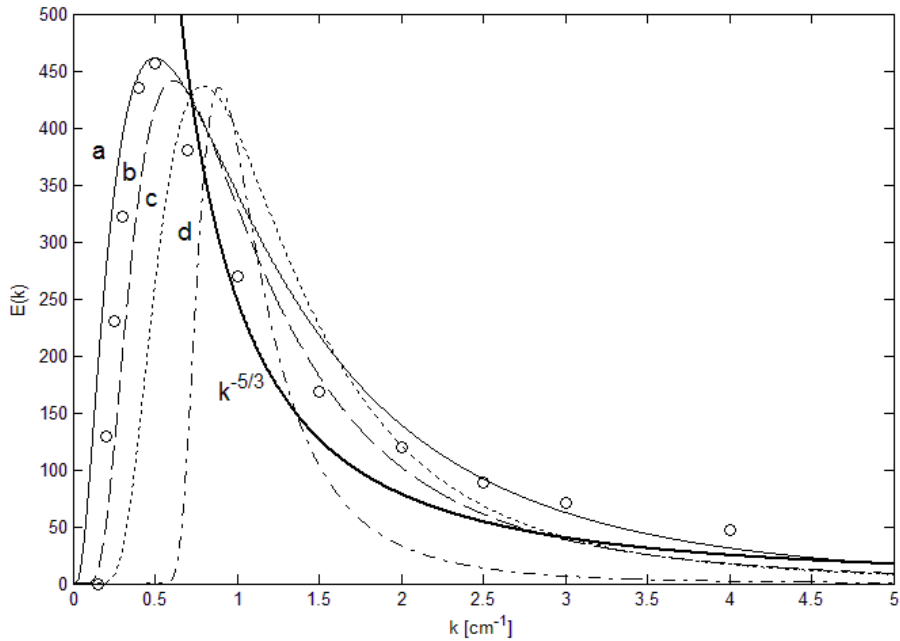

**Figure A1. Various u'(k) scaling used in Eq. 3 to generate the turbulence energy spectra: a. u'(k)=m-log(k); b. $k^{-1/2}$; c. $k^{-1/3}$; d. $k^{-3}$. Bold line is the Kolmogorov's $k^{-5/3}$ law in the inertial subrange. Symbols are data [9] at $Re_\lambda = 56$.**

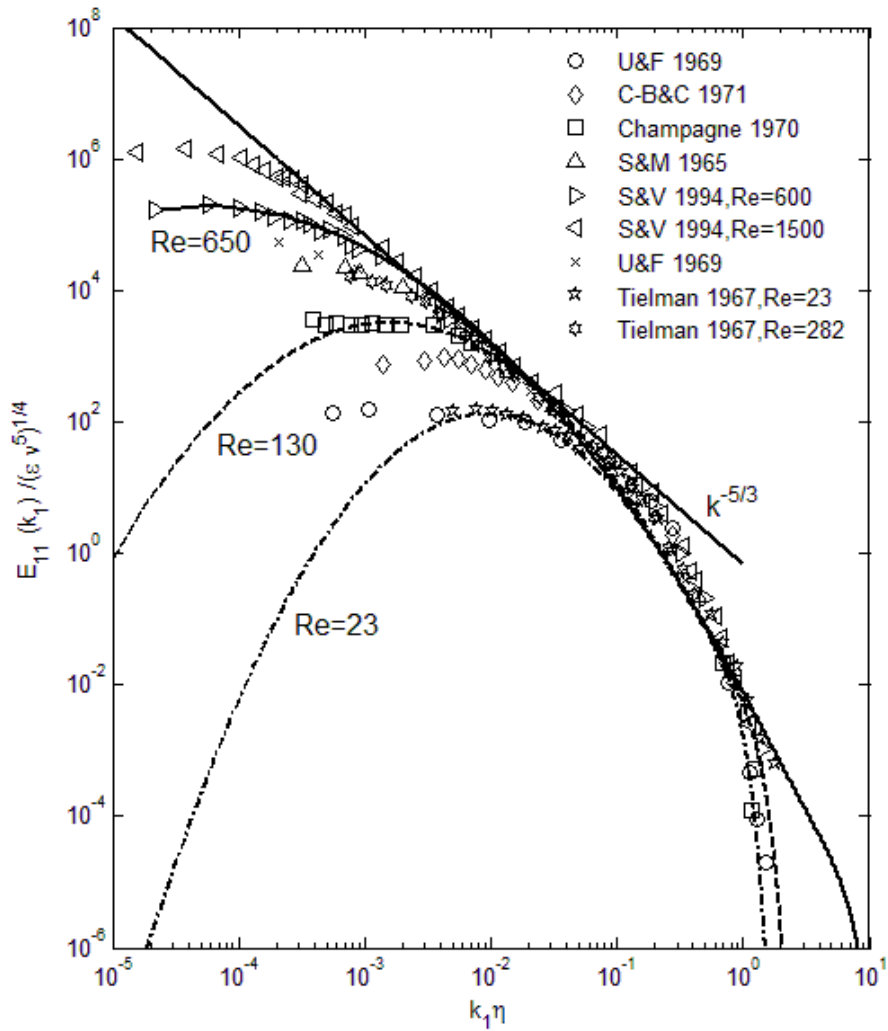

**Figure A2. Comparison of the turbulence energy spectra using Eq. 3 with experimental data. Various Reynolds numbers are tested from Re$_\lambda$ = 23 to 650. Kolmogorov scaling (k$^{-5/3}$) is also plotted. Symbols are data [9-15].**

In Figure 2, we can see that the viscous effect starts to be significant at high wavenumbers ($k\eta \sim 1$), where it causes a rapid decay toward zero energy. This is observable in some experimental data. We can further examine the effect of various constraints in the turbulence energy spectra. For small v, the deviation from the lognormal distribution is minimal. With increasing viscosity, the bend toward zero energy becomes more abrupt at high wavenumbers. The cut-off wavenumber for this bend does not appear to change appreciably with the viscosity, although at 2ν the effect of viscosity is transmitted to mid-range wavenumbers.

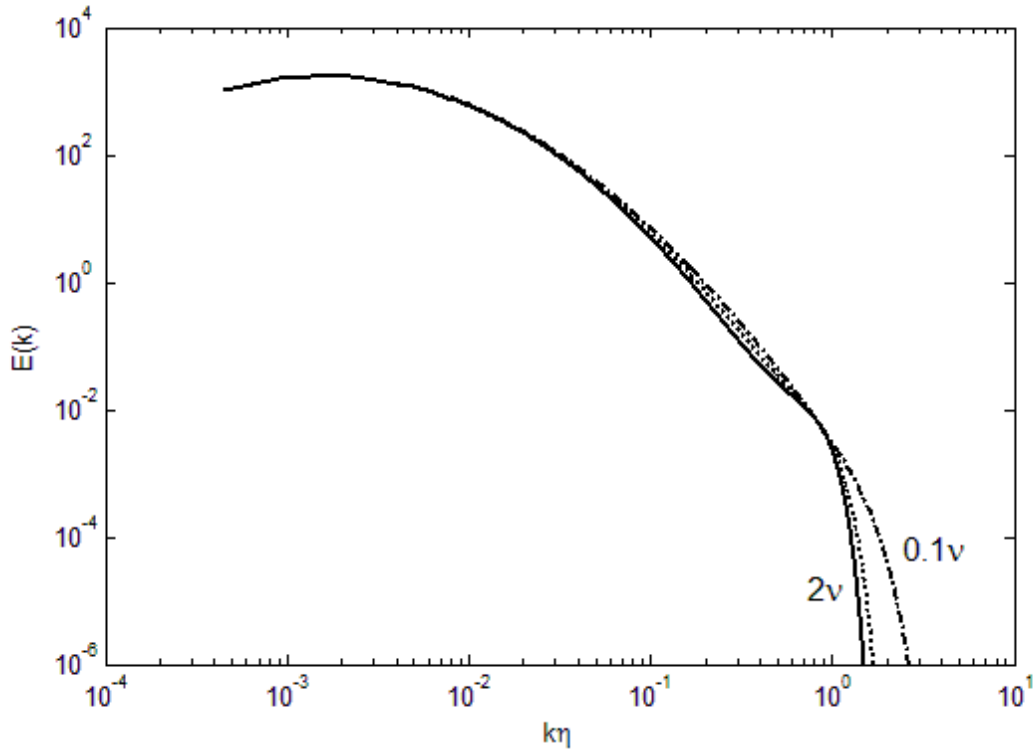

**Figure A3. Effect of viscosity on the turbulence energy spectra.**

The parameter, C2, is varied in Figure 4. $C_2$ is a "width" parameter specifying the range of length scales from say, Kolmogorov to the Taylor microscale. Due to the exponential nature of the distribution, a small change in $C_2$ leads to a large variations in the width of the distribution as shown in Figure 4. As discussed in a related work [6] and below, $C_2$ and all other parameters in Eq. 3 can be prescribed from known "information" about turbulence. For example, $C_2$ should scale with $\eta/\lambda$, or $Re_\lambda^{-1/2}$, meaning that at large Reynolds numbers $C_2$ is smaller leading to a broad energy spectra covering several orders of length scales.

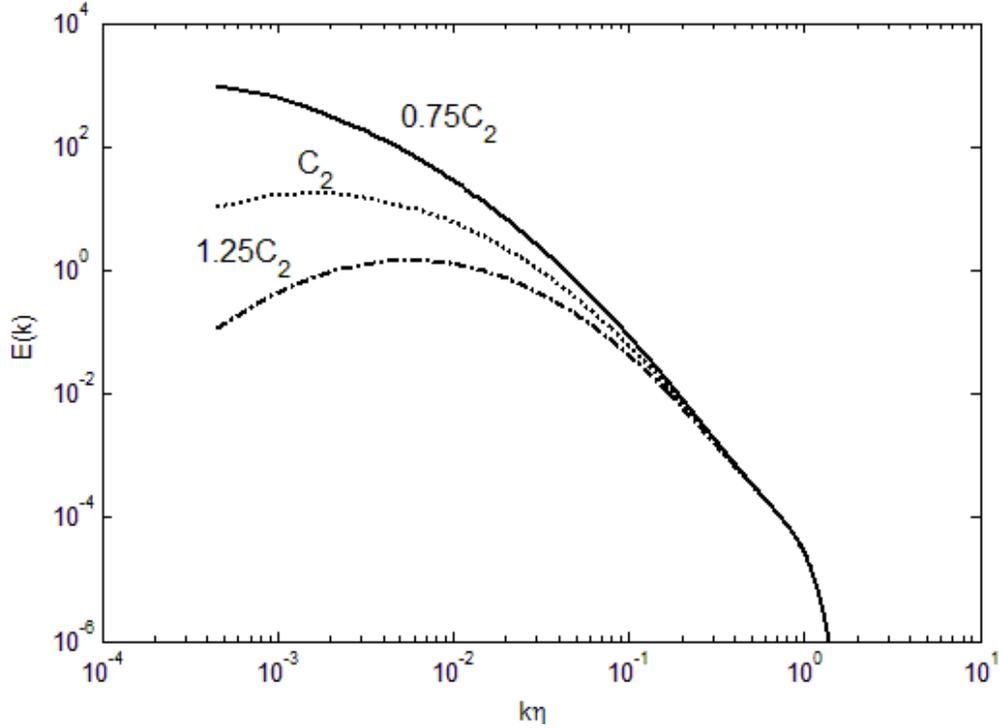

**Figure A4. Scaling of the turbulence energy spectra with variations in $C_2$ parameter.**

## APPENDIX B

For simple boundary layer flows, one form of Navier-Stokes equation is:

$$\frac{\partial(u^2)}{\partial x} + \frac{\partial(uv)}{\partial y} = -\frac{1}{\rho}\frac{dp}{dx} + \frac{1}{v}\frac{\partial^2 u}{\partial y^2} \qquad (A1)$$

The instantaneous velocities, u and v, are typically decomposed into the time mean (U, V) and fluctuating (u', v') components, u = U +u', and v = V + v', and numerical methods are applied to solve the so-called Reynolds-averaged Navier-Stokes equation (RANS). It is the cross-products of fluctuating components that lead to additional unknowns, the Reynolds stress (u'$^2$, u'v'), with insufficient number of equations to solve for them. This is known as the "closure problem", and recipes to obtain numerical solutions with modelled Reynolds stresses have become an industry called turbulence modeling.

Recently, we have shown that by applying the momentum balance to a coordinate frame moving at the mean velocity an alternate, Lagrangian prescription for the Reynolds stress can be obtained [14, 15]. This approach involves using a simple Galilean transform in Eq. A1: U+u' → u', and V+v' → v'. Under this transform, Eq. A1 becomes

$$\frac{d(u'v')}{dy} = -C_1 U \left[\frac{d(u'^2)}{dy} - \frac{1}{\rho}\frac{d|P|}{dy}\right] + \frac{1}{v}\frac{d^2 u'}{dy^2} \qquad (A2)$$

In Eq. A2, d/dx is replaced with $C_1 U d/dy$ to account for the displacement effect. This concept was derived from boundary layer and jet flows, but it appears to work for channel flows as well [14, 15]. Using Eq. A2, Reynolds stress can be directly computed, as validated in Figure A1, where the DNS data of Graham et al. [16] are used at $Re_\tau = 1000$.

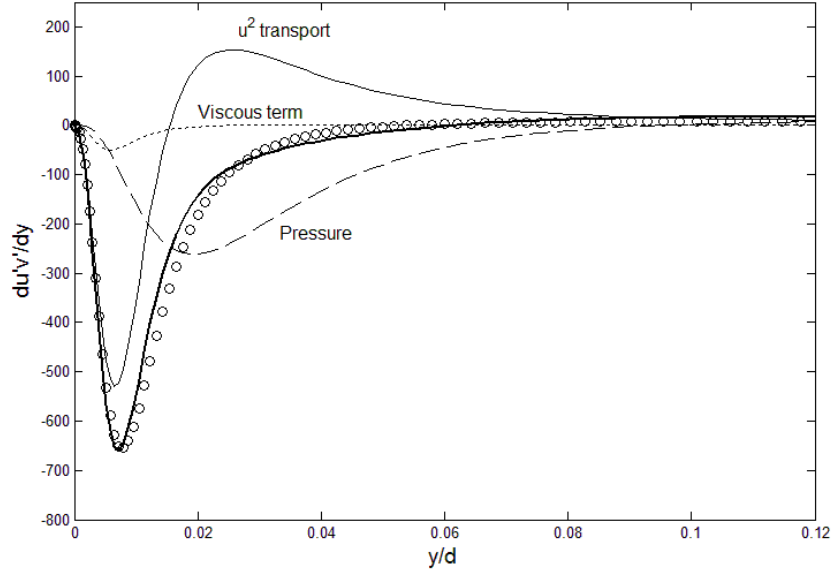

**Figure A1. Reynolds stress gradient budget. DNS channel flow data (circle symbol) for $Re_\tau = 1000$ [17] are used. Solid line is the RHS side of Eq. A2, with $u^2$-transport, pressure and the viscous terms combined.**

Eq. A2 is an expression that relates the off-diagonal Reynolds stress term, but we still need the primary diagonal component, $u'^2$. The Lagrangian formulation can be used again for transport of $u'^2$, which gives

$$C_2 U \frac{\partial (<u'^3>)}{\partial y} = -\frac{\partial (<u'^2 v'>)}{\partial y} + 2\nu_m \left(\frac{\partial u_{rms}'}{\partial y}\right)^2 \tag{A3}$$

27